\documentclass[aps,preprint]{revtex4}%
\usepackage{amsfonts}
\usepackage{amsmath}
\usepackage{amssymb}
\usepackage{graphicx}%
\begin{document}
\title{Microscopic Theory of the Photon Recoil of an Atom in a Dielectric}
\author{Hao Fu}
\author{P. R. Berman}
\affiliation{FOCUS, MCTP, and Physics Department, University of Michigan, Ann Arbor,
Michigan 48109-1040}
\pacs{PACS number}

\begin{abstract}
An atom recoils when it undergoes spontaneous decay. In this paper we present
a microscopic calculation of the recoil of a source atom imbedded in a
dielectric medium. We find that the source atom recoils with the canonical
photon momentum $n\hbar k_{0,}$ where $n$ is the index of refraction and
$\hbar k_{0}$ is the photon momentum calculated at the source atom atomic
frequency $\omega_{0}$. We also show explicitly how the energy is conserved
with the photon inside the medium.

\end{abstract}
\date{Dec 08 2005}
\maketitle

\section{Introduction}

The momentum of a photon in a dispersive medium has been considered by many
authors, due to its conceptual and practical importance. One of the issues is
how the momentum is conserved when a photon with momentum $\hbar\mathbf{k}$ in
the vacuum is scattered from an atom in the medium to a new momentum state
$\hbar\mathbf{k}^{\prime}$ in the vacuum. Should the momentum imparted on the
atom be the difference of the momenta in the vacuum $\hbar\mathbf{k}^{\prime
}-\hbar\mathbf{k}$ or the difference of the canonical photon momenta in the
medium $n\left(  \hbar\mathbf{k}^{\prime}-\hbar\mathbf{k}\right)  $
\cite{Milonni}? One might argue\cite{Ketterler} that, assuming the medium is
dilute, the atom is localized in the vacuum space between particles of the
medium, therefore the photon, before and after it hits the atom, travels in
the vacuum and momentum should be conserved in terms of the vacuum momenta. On
the other hand \cite{Haugen}, one can also argue that it is the macroscopic
field of the incident wave that induces and interacts with polarizations and
polarization currents of atoms in the medium and therefore the imparted
momentum should be the difference of the canonical photon momenta in the
medium. \ Experimentally, this issue has been studied in two systems. One
measures the recoil of a mirror immersed in a liquid when the light is
reflected from it \cite{Mirror}, and more recently \cite{Ketterler}, a
measurement of the recoil frequency of the Bose condensed $^{87}Rb$ using a
two-pulse Ramsey interferometer. \ Both experiments confirm that atoms recoil
according to canonical photon momenta.

Most theoretical studies related to this issue deal mainly with classical
fields. Some of the recent work by Loudon \cite{Loudon} and Nelson
\cite{Nelson} clarified some issues related to momentum in a dielectric from a
quantum and microscopic perspective. Milonni and Boyd \cite{Milonni} consider
a case where a source atom imbedded in the medium recoils due to its
spontaneous decay. They find that the source atom recoils according to
$n\hbar\omega_{0}/c$, where $\omega_{0}$ is the atomic frequency and $c$ is
the speed of light in the vacuum. Their calculation is based on a Heisenberg
Picture approach, where the operator expectation value, $\left\langle
P^{2}\right\rangle $ is calculated to be $n^{2}\hbar^{2}\omega_{0}^{2}/c^{2}$.
Here we present a similar calculation in the Schr\"{o}dinger Picture. The
calculation in the Schr\"{o}dinger Picture is particularly revealing because
it includes explicitly processes that are responsible for the modification of
the momentum imparted by the photon. As we show in the following, the photon
travelling in the medium experiences a series of scatterings from medium
atoms. Different scattering amplitudes interfere to shift the central
frequency of photons in the field. Since the source atom is coupled directly
to the field, by momentum conservation, the source atom recoils according to
the modified central frequency of photons. Our calculation is based on a
quantum field quantized in the free space, which allows a separate description
of the field and the medium, i.e. any wavelength and frequency of the field
are calculated unambiguously in the vacuum.

\section{Model and the Recoil Calculation}

The calculation is based on a model that we have used previously
\cite{First}\cite{Second}. The source atom, with finite mass $M,$ centered at
position $\left\langle \mathbf{R}\right\rangle =0$, has two internal levels,
whose frequency separation is denoted by $\omega_{0}$. The uniformly
distributed dielectric atoms have $J=0$ ground states and $J=1$ excited
states. The frequency separation of the ground and excited states is denoted
by $\omega$. We assume the mass of the medium atoms to be infinite, which
allows us to ignore recoil of the medium atoms. At $t=0$, the source atom is
excited to the $m=0$ excited state sublevel with center of mass momentum
$\left\langle P\right\rangle =0$, and $\left\langle \Delta P^{2}\right\rangle
\ll\left(  \hbar k_{0}\right)  ^{2}$, the dielectric atoms are all in their
ground states, and there are no photons in the field. The process we consider
one in which radiation emitted by the source atom is scattered by dielectric
atoms. The medium is modeled to be infinite, i.e. photons are always inside
the medium. The vaccum field amplitudes, the medium atoms' excited state
amplitudes, and the source atom center of mass motion is caculated as
$t\rightarrow\infty.$ It is assumed that the medium atoms are far detuned from
the source atom, $\omega\gg\omega_{0},$ and also $\omega_{0}\gg\gamma,$ where
$\gamma$ is the spontaneous decay rate of the source atom.

The free part of Hamiltonian describing such a system is
\begin{equation}
H_{0}=\frac{\hbar\omega_{0}}{2}\sigma_{z}+\sum_{j}\sum_{m=-1}^{1}\frac
{\hbar\omega}{2}\sigma_{z}^{(j)}(m)+\hbar\omega_{\mathbf{k}}a_{\mathbf{k}%
\lambda}^{\dagger}a_{\mathbf{k}\lambda}+\frac{\mathbf{P}^{2}}{2M}%
\end{equation}
where $\sigma_{z}=\left(  \left\vert 2\right\rangle \left\langle 2\right\vert
-\left\vert 1\right\rangle \left\langle 1\right\vert \right)  $, $\left\vert
2\right\rangle $ and $\left\vert 1\right\rangle $ are the $m=0$ excited and
$J=0$ ground state kets of the source atom, respectively, $\sigma_{z}%
^{(j)}(m)=$ $\left(  \left\vert m\right\rangle ^{(j)}\left\langle m\right\vert
-\left\vert g\right\rangle ^{(j)}\left\langle g\right\vert \right)  $ is the
population difference operator between excited state $\left\vert
J=1,m\right\rangle $ and ground state $\left\vert J=0,g\right\rangle $ \ of
dielectric atom $j$, and $a_{\mathbf{k}\lambda}$ is the annihilation operator
for a photon having momentum $\mathbf{k}$ and polarization $\lambda$. We have
also included a term describing the external motion of the source atom, where
$\mathbf{P}$ is the momentum operator for the source atom. A summation
convention is used, in which any repeated symbol on the right hand side of an
equation is summed over, unless it also appears on the left-hand side of the
equations. The interaction part that couples the field with the atoms is%

\begin{align}
V  &  =\hbar g_{\mathbf{k}}(\mathbf{\mu}_{0}\mathbf{\cdot\epsilon}%
_{\mathbf{k}}^{\left(  0\right)  }\sigma_{+}a_{\mathbf{k}}e^{i\mathbf{k}%
\cdot\mathbf{R}_{0}}-\mathbf{\mu}_{0}\mathbf{\cdot\epsilon}_{\mathbf{k}%
}^{\left(  0\right)  }a_{\mathbf{k}}^{\dagger}\sigma_{-}e^{-i\mathbf{k}%
\cdot\mathbf{R}_{0}})\nonumber\\
&  +\hbar g_{\mathbf{k\lambda}}^{\prime}\left[
\begin{array}
[c]{c}%
\mathbf{\mu}_{m}\mathbf{\cdot}\epsilon_{\mathbf{k}}^{\left(  \lambda\right)
}\sigma_{+}^{(j)}(m)\left(  a_{\mathbf{k}\lambda}e^{i\mathbf{k\cdot R}%
}-a_{\mathbf{k}\lambda}^{\dagger}e^{-i\mathbf{k\cdot R}}\right) \\
+\mathbf{\mu}_{m}^{\ast}\mathbf{\cdot}\epsilon_{\mathbf{k}}^{\left(
\lambda\right)  }\sigma_{-}^{(j)}(m)e^{-i\mathbf{k\cdot R}}\left(
a_{\mathbf{k}\lambda}^{\dagger}e^{-i\mathbf{k\cdot R}}-a_{\mathbf{k}\lambda
}e^{i\mathbf{k\cdot R}}\right)
\end{array}
\right]
\end{align}%
\begin{align}
g_{\mathbf{k}}  &  =-i\sqrt{\frac{\omega_{\mathbf{k}}}{2\hbar\epsilon_{0}V}%
}\label{coupling}\\
g_{\mathbf{k\lambda}}^{\prime}  &  =-i\sqrt{\frac{\omega_{\mathbf{k}}}%
{2\hbar\epsilon_{0}V}},
\end{align}
where $\sigma_{\pm}$ are raising and lowering operators for the source atom
and $\sigma_{\pm}^{(j)}(m)$ are raising and lowering operators between the
excited state $\left\vert J=1,m\right\rangle $ and the ground state
$\left\vert J=0,g\right\rangle $ of dielectric atom $j$, and $\mathbf{\mu}%
_{0}$ is the matrix element of the dipole operator for the source atom and
$\mathbf{\mu}_{m}$ is the matrix element of the medium atom between levels
$\left\vert J=1,m\right\rangle $ and $\left\vert J=0,g\right\rangle $. We have
taken the reduce matrix element, $\mu,$ of dipole moments of the source and
medium atoms to be equal to simplify our calculation. The polarization vectors
are defined as%
\begin{align}
\epsilon_{\mathbf{k}}^{(1)}  &  =\cos\theta_{\mathbf{k}}\cos\phi_{\mathbf{k}%
}\mathbf{\hat{x}}+\cos\theta_{\mathbf{k}}\sin\phi_{\mathbf{k}}\mathbf{\hat{y}%
}-\sin\theta_{\mathbf{k}}\mathbf{\hat{z}}\label{polarization}\\
\epsilon_{\mathbf{k}}^{(2)}  &  =-\sin\phi_{\mathbf{k}}\mathbf{\hat{x}}%
+\cos\phi_{\mathbf{k}}\mathbf{\hat{y}.} \label{Polarization}%
\end{align}
The source atom interacts only with the $z$ component of the vacuum field.
Since the medium atoms are far detuned from the source atom, we include
anti-rotating terms in the field-medium atoms' interaction. We have not
included such terms for the interaction Hamiltonian between the source atom
and the field because we have chosen the initial state to be the source atom
excited with no photon in the field.

Instead of writing amplitude equations for different states and then
identifying terms corresponding to contributions from different processes
\cite{Second}, we adopt a resolvent approach \cite{Tannoudji}, which allows us
to write amplitudes directly from diagrammatic representations of the
scattering processes. We want to calculate the recoil energy of the source
atom, which includes the contribution from three amplitudes: the source
recoiled atom with one photon in the field, the recoiled source atom with one
medium atom excited, and the recoiled source atom with both the field and a
medium atom excited. These amplitudes are represented by,
\begin{align}
b_{k}  &  =\left\langle 1,\mathbf{q};g;\mathbf{k}\right\vert U\left(
\infty\right)  \left\vert 2,0;g;0\right\rangle \label{bk}\\
b_{mj}  &  =\left\langle 1,\mathbf{q};m_{j};0\right\vert U\left(
\infty\right)  \left\vert 2,0;g;0\right\rangle \label{bmj}\\
b_{mjkk^{\prime}}  &  =\left\langle 1,\mathbf{q};m_{j};\mathbf{k}%
,\mathbf{k}^{\prime}\right\vert U\left(  \infty\right)  \left\vert
2,0;g;0\right\rangle , \label{bmkk}%
\end{align}
where the state of the whole system is labelled with source atom internal
states, source atom wave vector, medium atoms internal states, and quantum
field wave vectors. For example, $\left\vert 1,\mathbf{q};g;\mathbf{k}%
\right\rangle $ is a state of the source atom in the ground state $\left\vert
1\right\rangle $, with a momentum $\hbar\mathbf{q}$, the medium atoms in the
ground state $\left\vert g\right\rangle $, and a photon with a wave vector
$\mathbf{k}$ present. We use $\left\vert m_{j}\right\rangle $ to label the
state with a medium atom at position $R_{j}$ being excited to the $\left\vert
J=1,m\right\rangle $ state. The photon polarizations are not written
explicitly in the formulas. It should be understood, in the following
perturbative calculations, any intermedia states' photon polarzations are
summed over in amplitudes, while the final states' photon polarizations are
summed over in probabilities.

\begin{figure}
[ptb]
\begin{center}
\includegraphics[
height=2.5093in,
width=4.1165in
]%
{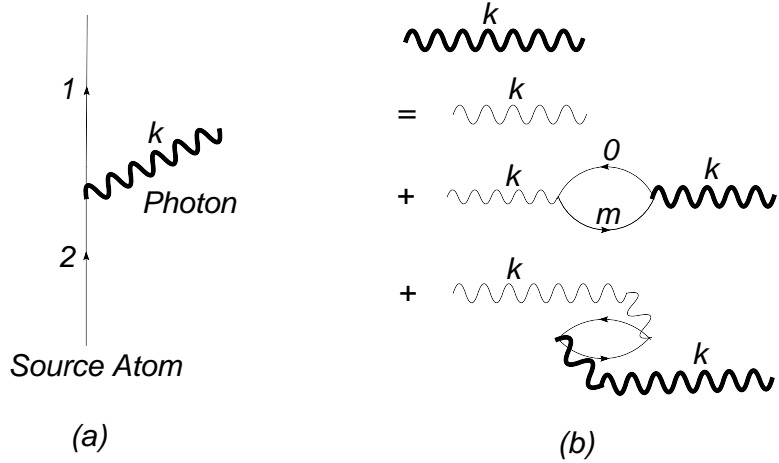}%
\caption{(a) shows that the excited source atom $\left\vert 2\right\rangle $
spontaneously decays by radiating a normalized photon with wave vector
$\mathbf{k}$ and goes to the ground state $\left\vert 1\right\rangle $ with
external momentum $-\hbar\mathbf{k}$. The thick wavy line corresponds to the
normalized photon propagator\ and striaght solid lines correspond to the atom
propagator. (b) shows the photon of wavevector $\mathbf{k}$ being scattered by
media atoms. We include both the rotating and anti-rotating contributions in
the intermedium states. Medium atoms can get excited to $J=1$ sublevel $m$ by
absorbing a photon $k$ and then go to the ground states by radiating photons.
They can also get excited by emitting photons and then deexcited by absorbing
other photons. In this diagram, the time propagation is from the left to the
right. Averaging over the positions of the medium atoms results in
conservation of the photon momenta before and after the scattering from medium
atoms. }%
\label{diagram}%
\end{center}
\end{figure}

The evolution operator can be expressed in terms of the retarded propagator
as
\begin{equation}
U\left(  \tau\right)  =-\frac{1}{2\pi i}\int_{-\infty}^{\infty}dE\exp\left(
-iE\tau/\hbar\right)  G\left(  E\right)  \label{evolution}%
\end{equation}
and the resolvent operator, $G\left(  E+i0\right)  $, is defined as
\[
G\left(  E\right)  =\frac{1}{E+i0-\hat{H}}%
\]
where $H$ is the total Hamiltonian, $E$ is the energy of the system, and the
infinite small imaginary $i0$ prescription is used to yield retarded
propagation. In order to carry out a perturbative calculation in orders of the
interaction $\hat{V},$ we recast the resolvant operator in the following exact
form%
\begin{equation}
G\left(  E\right)  =G^{\left(  0\right)  }\left(  E\right)  +G^{\left(
0\right)  }\left(  E\right)  \hat{V}G\left(  E\right)  \label{reex}%
\end{equation}
Here $G^{\left(  0\right)  }\left(  E\right)  $ is the zeroth order
propagator
\[
G^{\left(  0\right)  }\left(  E\right)  =\frac{1}{E+i0-\hat{H}_{0}}.
\]
The processes that we want to take into account are shown in Fig(\ref{diagram}%
).
Let's focus on the calculation of $b_{k}$ first,
\begin{equation}
b_{k}=-\frac{1}{2\pi i}\lim_{\tau\rightarrow\infty}\int_{-\infty}^{\infty
}dE\exp\left(  -iE\tau/\hbar\right)  \left\langle 1,\mathbf{q};g_{m}%
;\mathbf{k}\right\vert G\left(  E\right)  \left\vert 2,0;g_{m};0\right\rangle
,
\end{equation}
which requires a calculation of the matrix element of the operator. According
to equation (\ref{reex}), the resolvant operator can be expanded to first
order in $\hat{V}$ as,
\begin{align}
&  \left\langle 1,\mathbf{q};g_{m};\mathbf{k}\right\vert G^{\left(  1\right)
}\left\vert 2,0;g_{m};0\right\rangle \nonumber\\
&  =\left\langle 1,\mathbf{q};g_{m};\mathbf{k}\right\vert G^{\left(  0\right)
}\left(  E\right)  \left\vert 1,\mathbf{q};g_{m};\mathbf{k}\right\rangle
\left\langle 1,\mathbf{q};g_{m};\mathbf{k}\right\vert \hat{V}\left\vert
2,0;g_{m};0\right\rangle \left\langle 2,0;g_{m};0\right\vert G^{\left(
0\right)  }\left(  E\right)  \left\vert 2,0;g_{m};0\right\rangle \nonumber\\
&  =\hbar g_{k}^{\ast}\delta_{\mathbf{q},-\mathbf{k}}\left(  \mu_{m}^{\ast
}\cdot\epsilon_{k}\right)  \frac{1}{E-\hbar\omega_{0}+i\hbar n\gamma}\frac
{1}{E+i0-\hbar\omega_{q}^{r}-\hbar\omega_{k}}\mathbf{W}_{m,0}. \label{zero}%
\end{align}
We neglect the possibility that the photon can be scattered by medium atoms in
this order. In the calculation, the source atom recoil momentum equals the
inverse of the photon momentum, $\mathbf{q=-k}$ (conservation of momentum),
resulting from evaluating the matrix element $\left\langle \mathbf{p}%
=\hbar\mathbf{q}\right\vert e^{i\mathbf{k\cdot R}}\left\vert \mathbf{p}%
=0\right\rangle $. We have rearranged the order in the last line of the above
formula (\ref{zero}) so that the first part,
\begin{equation}
\hbar g_{k}^{\ast}\delta_{\mathbf{q},-\mathbf{k}}\left(  \mu_{m}^{\ast}%
\cdot\epsilon_{k}^{\left(  \lambda\right)  }\right)  \frac{1}{E-\hbar
\omega_{0}+i\hbar n\gamma},
\end{equation}
describes the decay of the source atom in the medium. The imaginary part in
the dominator $i\hbar n\gamma$ is added to represent spontaneous decay
\cite{Tannoudji}. The modification of $\gamma$ in the vacume to $n\gamma$ in
the medium is related to the process that photons scattered by the media atoms
are reabsorbed by the source atom, as shown in papers \cite{First}%
\cite{Second}. The second part of Equation(\ref{zero}),
\begin{equation}
\frac{1}{E+i0-\hbar\omega_{q}^{r}-\hbar\omega_{k}}\mathbf{W}_{m,0},
\label{php0}%
\end{equation}
describes the propagation of the photon, where $\omega_{q}^{r}=\hbar q^{2}/2M$
is the recoil frequency associated with the emission of a phone with
wavevector $q$ and $\mathbf{W}_{m,0}$ is the transverse polarization tensor of
the photon defined as
\begin{align}
\mathbf{W}_{m,m^{\prime}}  &  =\left(  \mathbf{\mu}_{m}\cdot\mathbf{\epsilon
}_{k}^{\left(  \lambda\right)  }\right)  \left(  \mathbf{\mu}_{m^{\prime}%
}\cdot\mathbf{\epsilon}_{k}^{\left(  \lambda\right)  }\right) \nonumber\\
&  =\left(  \mathbf{\epsilon}_{k}^{\left(  \lambda\right)  }\otimes
\mathbf{\epsilon}_{k}^{\left(  \lambda\right)  }\right)  _{m,m^{\prime}%
}=\left(  1-\hat{k}\otimes\hat{k}\right)  _{m,m^{\prime}}%
\end{align}
where two relationships
\[
\mathbf{\mu}_{m}^{\ast}\mathbf{\mu}_{m}=\mathbf{1}%
\]
and
\begin{equation}
\mathbf{\mu}_{0}^{\ast}\cdot\mathbf{\epsilon}_{k}=\left(  \mathbf{\epsilon
}_{k}\cdot\mathbf{\mu}_{m}^{\ast}\right)  \mathbf{W}_{m,0}%
\end{equation}
have been used.

In order to carry the calculation of the matrix element of the resolvant
operator, $\left\langle 1,\mathbf{q};g_{m};\mathbf{k}\right\vert G\left\vert
2,0;g_{m};0\right\rangle ,$ to higher order, it is neccessary to consider
processes that the photon is scattered by medium atoms. Including these
processes modifies the photon propagator(\ref{php0}). For a dilute medium with
$N\lambda_{0}^{3}\ll1,$ where $N$ is the density of the medium and
$\lambda_{0}=2\pi c/\omega_{0}$ is the photon wavelength, it is appropriate to
make use of the independent scattering approximation. Namely, we need to
include only contributions from processes shown in Fig[\ref{diagram}b], i.e.
ladder diagrams, which amount to a self energy insertion \cite{Tannoudji} to
the photon propagator(\ref{php0}) in the dominator as the following%
\begin{align}
\sum &  =\sum_{j,m,k^{\prime}}\left\langle 1,\mathbf{q};g;\mathbf{k}%
\right\vert V\nonumber\\
&  \times\left(
\begin{array}
[c]{c}%
\left\vert 1,\mathbf{q};m_{j};\mathbf{k},\mathbf{k}^{\prime}\right\rangle
\left\langle 1,\mathbf{q};m_{j};\mathbf{k},\mathbf{k}^{\prime}\right\vert
G^{\left(  0\right)  }\left(  E\right)  \left\vert 1,\mathbf{q};m_{j}%
;\mathbf{k},\mathbf{k}^{\prime}\right\rangle \left\langle 1,\mathbf{q}%
;m_{j};\mathbf{k},\mathbf{k}^{\prime}\right\vert \\
+\left\vert 1,\mathbf{q};m_{j};0\right\rangle \left\langle 1,\mathbf{q}%
;m_{j};0\right\vert G^{\left(  0\right)  }\left(  E\right)  \left\vert
1,\mathbf{q};m_{j};0\right\rangle \left\langle 1,\mathbf{q};m_{j}%
;0\right\vert
\end{array}
\right) \nonumber\\
&  \times V\left\vert 1,\mathbf{q};g;\mathbf{k}^{\prime}\right\rangle
\end{align}
Using the prescription $\sum_{R_{j}}\rightarrow N\int d\mathbf{R}$ and
$\sum_{k^{\prime}}\rightarrow\frac{V}{\left(  2\pi\right)  ^{3}}\int
d\mathbf{k}^{\prime}$ we can change sums to integrals. The integration over
$R_{j}$ gives $\delta\left(  \mathbf{k}-\mathbf{k}^{\prime}\right)  $ as a
result of the translational invariance of the medium. The integration over
$\mathbf{k}^{\prime}$ picks up contributions only at $\mathbf{k}$ and the self
energy can be written as
\begin{equation}
\sum=\hbar^{2}NV\left\vert g_{k}\right\vert ^{2}\mu^{2}\left[  \frac
{1}{E-\hbar\omega_{k}^{r}-\hbar\omega}+\frac{1}{E-\hbar\omega_{k}^{r}%
-\hbar\omega-2\hbar\omega_{k}}\right]  \label{self}%
\end{equation}
Spontaneous decays of the medium atoms are ignored because medium atoms are
far detuned from the source atom. The normalized photon propagator, with the
above self energy modification, is
\begin{equation}
\frac{1}{E+i0-\hbar\omega_{k}^{r}-\hbar\omega_{k}-\sum}\mathbf{W=}\frac
{1}{E-\hbar\omega_{k}^{r}-\left[  1-\frac{1}{2}N\alpha\left(  E\right)
\right]  \hbar\omega_{k}}\mathbf{W,} \label{np}%
\end{equation}
where the polarizability is defined as%
\begin{equation}
\alpha\left(  E\right)  =-\frac{4\pi\mu^{2}}{\Delta\left(  E\right)  }
\label{alpha}%
\end{equation}
and the detunning
\begin{equation}
\frac{1}{\Delta\left(  E\right)  }\equiv\frac{1}{E-\hbar\omega_{k}^{r}%
-\hbar\omega}+\frac{1}{E-\hbar\omega_{k}^{r}-\hbar\omega-2\hbar\omega_{k}}%
\end{equation}
\ The self energy insertion(\ref{self}) brings a correction of order $N\alpha$
to the denominator of the photon propagator, which cannot be obtained from any
finite order calculation. Substituting this normalized photon
propagator(\ref{np}) back to the first order formula(\ref{zero}) and using the
fact that $\mathbf{W}^{2}=\mathbf{W}$, one fines the relevent reolvant matrix
element to be%
\begin{align}
&  \left\langle 1,-\hbar\mathbf{k};g_{m};\mathbf{k};0\right\vert G\left(
E+i0\right)  \left\vert 2,0;g_{m};0\right\rangle \nonumber\\
&  =\hbar g_{k}^{\ast}\left(  \mu_{0}^{\ast}\cdot\epsilon_{k}\right)  \frac
{1}{E-\hbar\omega_{0}+in\hbar\gamma}\frac{1}{E-\hbar\omega_{k}^{r}-\left(
1-\frac{1}{2}N\alpha\right)  \hbar\omega_{k}}%
\end{align}
To find the transition amplitude in Equation(\ref{bk}), we need to integrate
over $E$ according to equation (\ref{evolution}). The integration includes
contributions from two poles, one at $E=\hbar\omega_{k}^{r}+\left(  1-\frac
{1}{2}N\alpha\right)  \hbar\omega_{k}$, and one\ at $E=\hbar\omega_{0}-i\hbar
n\gamma$. In the limit $\gamma\tau\gg1$, only the first pole contributes,
since the second one has a finite imaginary part and its contribution decays
away as $e^{-n\gamma\tau}$. The amplitude of finding a photon with wave number
$k$ and source atom with momentum $-\hbar\mathbf{k}$ is then%
\begin{equation}
b_{k}=\lim_{\tau\rightarrow\infty}g_{k}^{\ast}\left(  \mu_{0}^{\ast}%
\cdot\epsilon_{k}\right)  \frac{\exp\left[  -i\left(  \hbar\omega_{k}%
^{r}+\left(  1-\frac{1}{2}N\alpha\right)  \omega_{k}\right)  \tau\right]
}{\omega_{k}^{r}+\left(  1-\frac{1}{2}N\alpha\right)  \omega_{k}-\omega
_{0}+in\gamma} \label{bk1}%
\end{equation}
and the corresponding probability is \cite{Note1}
\begin{equation}
\left\vert b_{k}\right\vert ^{2}=\left\vert g_{k}\right\vert ^{2}\left\vert
\mu_{0}\cdot\epsilon_{k}\right\vert ^{2}\frac{1}{\left[  \omega_{k}/n-\left(
\omega_{0}-n^{2}\omega_{0}^{r}\right)  \right]  ^{2}+n^{2}\gamma^{2}}%
\end{equation}
with $\omega_{0}^{r}=\hbar\omega_{0}^{2}/2Mc^{2}.$ We have identified
$n=1+\frac{1}{2}N\alpha$ in the limit of small $N\alpha$ \cite{Note2}. The
photon frequency centers around $n\left(  \omega_{0}-n^{2}\hbar\omega_{0}%
^{r}\right)  .$ The total probability of finding a photon in the field is just
the sum over all the $k^{\prime}s$%
\begin{equation}
\sum_{k}\left\vert b_{k}\right\vert ^{2}=\frac{4\pi}{3}\frac{2\mu^{2}}{\left(
2\pi\right)  ^{2}\hbar c^{3}}\int d\omega_{k}\frac{\omega_{k}^{3}}{\left[
\omega_{k}/n-\left(  \omega_{0}-n^{2}\omega_{0}^{r}\right)  \right]
^{2}+n^{2}\gamma^{2}}=1 \label{one}%
\end{equation}
Note that in working out the integration, we have used the Wigner-Weisskopf
approximation. We extend the integration of $\omega_{k}$ to start from minus
infinity\cite{Paul}, and take into account only the contribution from the pole
at $n\left(  \omega_{0}-n^{2}\omega_{0}^{r}+in\gamma\right)  $\cite{Note4}.
Here the decay rate $\gamma$ should be evaluated at the photon frequency and
it is given by $\gamma=2\mu^{2}\omega_{k}^{3}/3\hbar c^{3}.$

\begin{figure}
[ptb]
\begin{center}
\includegraphics[
height=2.6938in,
width=3.5799in
]%
{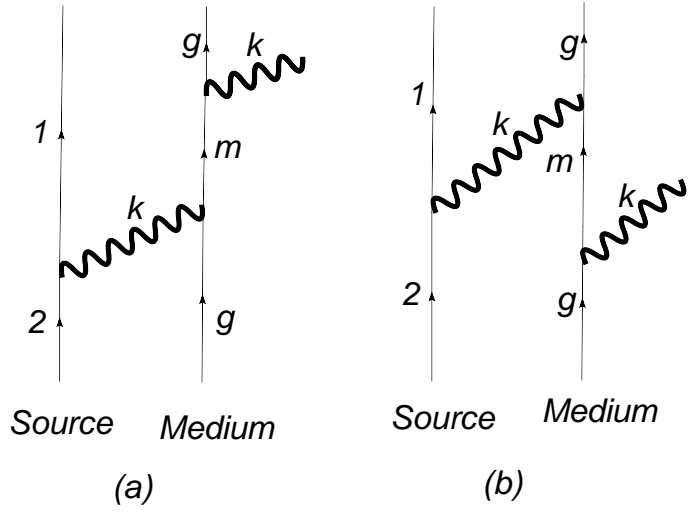}%
\caption{Medium atoms are excited by emitting\ or absorbing normalized
photons. Figure (a) corresponds to a rotating situation while figure (b)
corresponds to an anti-rotating situation.}%
\label{ex}%
\end{center}
\end{figure}

Other relevent amplitudes, indicated in Fig[\ref{ex}], can be calculated by
the same technique and now consider the processes shown in Fig[\ref{ex}].
Making use of the normalized photon propagator(\ref{np}), we find the
amplitude for a medium atom located at $R_{j},$ being excited to a sublevel
$m,$ without any photon present to be
\begin{align}
b_{m_{j}}  &  =\lim_{\tau\rightarrow\infty}\sum_{k}\left\vert g_{k}\right\vert
^{2}\mu^{2}e^{i\mathbf{k\cdot R}_{j}}\frac{\exp\left[  -i\left(  \hbar
\omega_{k}^{r}+\left(  1-\frac{1}{2}N\alpha\right)  \omega_{k}\right)
\tau\right]  }{\omega_{k}^{r}+\left(  1-\frac{1}{2}N\alpha\right)  \omega
_{k}-\omega_{0}+in\gamma}\\
&  \times\frac{1}{\left(  1-\frac{1}{2}N\alpha\right)  \omega_{k}-\omega
}\mathbf{W}_{m,0}%
\end{align}
It proves more useful to work in momentum space, since one can take advantage
of \ the translational invariance of the problem. We label the corresponding
momentum state amplitude as $b_{m}\left(  k\right)  $%
\begin{align}
b_{m}\left(  k\right)   &  =\lim_{\tau\rightarrow\infty}\sqrt{NV}\mu
^{2}\left\vert g_{k}\right\vert ^{2}\frac{\exp\left[  -i\left(  \hbar
\omega_{k}^{r}+\left(  1-\frac{1}{2}N\alpha\right)  \omega_{k}\right)
\tau\right]  }{\omega_{k}^{r}+\left(  1-\frac{1}{2}N\alpha\right)  \omega
_{k}-\omega_{0}+in\gamma}\nonumber\\
&  \times\frac{1}{\left(  1-\frac{1}{2}N\alpha\right)  \omega_{k}-\omega
}\mathbf{W}_{m,0}%
\end{align}
The probability of finding medium atoms excited is
\begin{align}
\sum_{m,k}\left\vert b_{m}\left(  k\right)  \right\vert ^{2}  &  =NV\mu
^{4}\sum_{k}\frac{\left\vert g_{k}\right\vert ^{4}}{\left[  \omega
_{k}/n-\left(  \omega_{0}-n^{2}\omega_{0}^{r}\right)  \right]  ^{2}%
+n^{2}\gamma^{2}}\nonumber\\
&  \times\left[  \frac{1}{\left(  1-\frac{1}{2}N\alpha\right)  \omega
_{k}-\omega}\right]  ^{2}\mathbf{W}_{0,0} \label{pm}%
\end{align}
The amplitude of the anti-rotating excitation, a medium atom excited with two
photons present can be shown to be
\begin{align}
b_{mkk}\left(  k\right)   &  =\lim_{t\rightarrow\infty}\sqrt{NV}\mu
^{2}\left\vert g_{k}\right\vert ^{2}\frac{\exp\left[  -i\left(  \hbar
\omega_{k}^{r}+\left(  1-\frac{1}{2}N\alpha\right)  \omega_{k}\right)
t\right]  }{\omega_{k}^{r}+\left(  1-\frac{1}{2}N\alpha\right)  \omega
_{k}-\omega_{0}+in\gamma}\nonumber\\
&  \times\frac{1}{\left(  1-\frac{1}{2}N\alpha\right)  \omega_{k}%
-\omega-2\omega_{k}}\mathbf{W}_{m,0}%
\end{align}
and the corresponding probability, summing over different magnetic sublevels,
\begin{align}
\sum_{m,k}\left\vert b_{mkk}\left(  k\right)  \right\vert ^{2}  &  =NV\mu
^{4}\sum_{k}\frac{\left\vert g_{k}\right\vert ^{4}}{\left[  \omega
_{k}/n-\left(  \omega_{0}-n^{2}\omega_{0}^{r}\right)  \right]  ^{2}%
+n^{2}\gamma^{2}}\nonumber\\
&  \times\left[  \frac{1}{\left(  1-\frac{1}{2}N\alpha\right)  \omega
_{k}-\omega-2\omega_{k}}\right]  ^{2}\mathbf{W}_{0,0} \label{pmkk}%
\end{align}
The probability $\sum_{m,k}\left\vert b_{m}\left(  k\right)  \right\vert ^{2}$
and $\sum_{m,k}\left\vert b_{mkk}\left(  k\right)  \right\vert ^{2}$ are of
order $N\alpha\omega_{k}/\omega,$ which are negligible with $\omega_{k}%
\ll\omega$ and $N\alpha\ll1.$ By now we have shown all the nonvanishing
amplitudes of the system at $\tau\rightarrow\infty$. The total probability,
sum of equations (\ref{one})(\ref{pm})(\ref{pmkk}), is found to be one to the
order of $N\alpha.$

The average recoil energy of source atom is calculated solely from the
amplitude $b_{k}$ (\ref{bk1}) as
\begin{equation}
\left\langle \frac{\hbar^{2}k^{2}}{2M}\right\rangle =\sum_{k}\frac{\hbar
^{2}k^{2}}{2M}\left\vert b_{k}\right\vert ^{2}=n^{2}\hbar\omega_{0}^{r}%
\end{equation}
This result shows that the source atom recoils according to the canonical
photon momentum $nk_{0}$. This modification of the source atom recoil is
directly related to the fact that the photon in the medium centers at a
frequency $n\omega_{0}+O\left(  \hbar\omega_{0}^{2}/Mc^{2}\right)  $ instead
of $\omega_{0}+O\left(  \hbar\omega_{0}^{2}/Mc^{2}\right)  $, as is shown in
formula (\ref{bk1}). In the spontaneous decay of the source atom, momentum of
the source atom plus the field is conserved. This modification of the center
freqency of the photon therefore results in a modification of the source atom
recoil. However, this shift in the photon frequency seems strange, because we
expect the frequency to center around the atomic frequency $\omega_{0}$ from a
energy conservation consideration. In the following, we give a detailed
analysis of the energy conservation to order of the shifted frequency, namely,
$n\hbar\omega_{0}-\hbar\omega_{0}=\frac{1}{2}N\alpha\hbar\omega_{0}$.

Before we proceed to show energy conservation, it is necessary to note that
the source atom excited at $t=0$ is not an eigenstate of the system. However,
a discussion of energy conservation is meaningful in the context of average
energy being conserved.

In order to find all the other forms of energy besides the photonic
excitation, we should note that as the photon moves along in the medium,
medium atoms inside the sphere of $R=ct$ are excited, according to
$e^{-\gamma\left(  ct-R\right)  }$. Even the excitation probability is of
order of magnitude $N\alpha\omega_{k}/\omega$, as far as the energy is
concerned, this produces a correction of order $N\alpha\omega_{k}$. Including
this contribution to the energy, and the interaction energy of the medium
atoms with the field, we should be able to recover the conservation of energy.
In the following we work out explicitly energies associated with different
excitations in the system and show that the total average energy is conserved.

The first part of the energy corresponding to the field excitation is%
\begin{equation}
\sum_{k}\left\vert b_{k}\right\vert ^{2}\hbar\omega_{k}=\left(  1+\frac{1}%
{2}N\alpha\right)  \hbar\omega_{0}%
\end{equation}
The one associated with the medium atoms excitations is
\begin{align}
&  \sum_{m,k}\left\vert b_{mk}\right\vert ^{2}\hbar\omega+\sum_{m,k}\left\vert
b_{mkk}\right\vert ^{2}\left(  \hbar\omega+2\hbar\omega_{k}\right) \nonumber\\
&  =-NV\sum_{k}\hbar^{4}\left\vert g_{k}\right\vert ^{4}\mathbf{W}_{00}%
\frac{1}{\left(  \hbar\omega_{k}/n-\hbar\omega_{0}\right)  ^{2}+\gamma^{2}%
}\frac{1}{\Delta^{\prime}}%
\end{align}
where
\[
\frac{1}{\Delta^{\prime}}\equiv-\hbar\omega_{m}\left\{  \left[  \frac
{1}{\left(  1-\frac{1}{2}N\alpha\right)  \hbar\omega_{k}-\hbar\omega}\right]
^{2}+\left[  \frac{1}{\left(  1+\frac{1}{2}N\alpha\right)  \hbar\omega
_{k}+\hbar\omega}\right]  ^{2}\right\}
\]
Here the difference in the energies of state $b_{mk}$ and $b_{mkk}$ can be
ignored, because it amounts to a correction in the order $N\alpha\omega
_{k}/\omega$. To leading order of $\omega_{k}/\omega_{m}$, we find that
\begin{equation}
\frac{1}{\Delta^{\prime}}-\frac{1}{\Delta\left(  E\right)  }=-\frac{1}%
{\hbar\omega}N\alpha\frac{\omega_{k}}{\omega}%
\end{equation}
which enables us to neglect the difference between $\frac{1}{\Delta^{\prime}}$
and $\frac{1}{\Delta\left(  E\right)  }.$ Here the detuning $\Delta\left(
E\right)  $ is evaluated at the pole of the photon propagator $E=\left(
1-\frac{1}{2}N\alpha\right)  \hbar\omega_{k}$. Making use of the definition
(\ref{alpha}), one finds that the energy associated with the atomic
excitations is $\frac{1}{2}N\alpha\hbar\omega_{0}$. The third part of the
energy, the interaction energy between excited medium atoms and the field can
be calculated as,
\begin{align}
\left\langle V\right\rangle  &  =\sum_{m,k}NV\hbar g_{k}\left(  b_{k}^{\ast
}b_{m}+b_{k}^{\ast}b_{mkk}\right)  +h.c.\nonumber\\
&  =\sum_{k}-N\alpha\left(  \omega_{k}\right)  \hbar\omega_{k}\frac{\left\vert
\hbar g_{k}\right\vert ^{2}}{\left(  \hbar\omega_{k}/n-\hbar\omega_{0}%
+n^{2}\hbar\omega_{k}^{r}\right)  ^{2}+n^{2}\hbar^{2}\gamma^{2}}\nonumber\\
&  =-N\alpha\left(  \omega_{k}\right)  \hbar\omega_{k}%
\end{align}
adding up all of these contributions, we have%
\begin{equation}
\sum_{k}\left\vert b_{k}\right\vert ^{2}\hbar\omega_{k}+\sum_{m,k}\left(
\left\vert b_{mk}\right\vert ^{2}+\left\vert b_{mkk}\right\vert ^{2}\right)
\hbar\omega+\left\langle V\right\rangle =\hbar\omega_{0}%
\end{equation}
showing that the frequency shift of the photon in the medium is compensated by
the excitation energy of medium atoms and the interaction energy of excited
atoms with the field.

\section{Discussion}

\bigskip In this paper we have shown that the source atom recoils according to
$n\hbar k_{0}$, which agrees with the previous theoretical and experimental
results \cite{Ketterler}\cite{Mirror}\cite{Milonni}. This modification of the
photon recoil arises in our calculation as a result of the interference of the
different scattering amplitudes of the photon. As has been shown in the
calculation, while the center frequency of the soure atom is at $\omega_{0}$,
only frequencies centered at $n\omega_{0}$ experience contructive
interference. This is very much similar to the case of a source atom radiating
in a cavity with the cavity frequency instead of the atomic frequency. From a
quantum point of view, the source atom decays because it radiates and
reabsorbs virtual photons. Such a process introduces a finite self energy
whose real part gives the level shift and whose imaginary part gives the decay
of the atomic excitation \cite{First}\cite{Second}\cite{Flei}. This process,
though not included explicitly in our calculation, is the only way that the
source atom "knows about" the enviroment(vacuum, cavity or dielectric medium)
in which it locates. In the cavity, the virtual photon can interfere with
those reflected from the cavity walls and constructive interference occurs
only at the cavity freqency. In the dielectric medium the virtual photons
radiated by the source atom can be scattered by medium atoms and reabsorbed by
the source atom during the time $t\preceq1/\gamma$. Different scattering
amplitudes interfere to shift the real radiating frequency to $n\omega_{0}$.
This is a different effect from the level shift, the real part of the source
atom self energy, due to the interaction of the source atom with the
enviroment. In our particular example of the source atom located in a dilute
medium, the shift due to interacting is of the order $N\alpha\gamma$ while the
shift due to interference is of the order $N\alpha\omega_{0}.$

\bigskip\ An alternative explaination can be put forward in terms of eigen
excitations of the system. Actually, if we consider the interaction of the
field with the medium atom, the eigenmode of the system is neither medium
atoms being excited nor a photon present, but a superposition of these two
type of excitations, or a polariton\cite{Polariton}. The source atom decays by
radiating polaritons instead of photons. For the energy to be conserved, the
polariton energy plus the recoil energy should equal to the initial average
energy $\hbar\omega_{0}$. As we have shown, the photon carries only part of
the energy of the eigen excitation. On the other hand, the momentum carried by
the medium atoms is negligible, the polariton momentum is just the photon
momentum. When we require energy and momentum conservation for the radiating
process, the only possibility is the source atom to recoil according to
$n\hbar k_{0}$.

\begin{acknowledgments}
We thank P. W. Milonni for a discussion concerning the conservation of energy.
This research is supported by the National Science Foundation under Grants No.
PHY-0244841 and the FOCUS Center Grant.
\end{acknowledgments}

\end{document}